\documentclass[a4paper]{PoS}
\usepackage{amsmath}
\usepackage{amssymb}
\usepackage{epsfig}
\usepackage{euscript}
\def\mathswitchr#1{\relax\ifmmode{\mathrm{#1}}\else$\mathrm{#1}$\fi}

\newcommand {\pslash}{\hbox{$\not\hbox{\kern-2.3pt $p$}$}}

\newcommand{\FYFS}{F_{\mathrm{YFS}}}
\usepackage{cite}

\usepackage{epic}

\def\alf1{ {\alpha\over\pi} }
\def\rQCED{{\rm QCED}}

\title{Interplay of IR-Improved DGLAP-CS Theory and NLO Parton Shower MC Precision}

\ShortTitle{Interplay of IR-Improved DGLAP-CS Theory and NLO Parton Shower MC Precision}

\author{\speaker{B.F.L. Ward}\\
        Baylor University\\
        E-mail: \email{bfl\_ward@baylor.edu}}

\author{ S.K. Majhi%
         \thanks{Work supported by 
grant Pool No. 8545-A, CSIR, IN.}\\
      Indian Association for the Cultivation of Science\\
        E-mail: \email{tpskm@iacs.res.in}}
\author{ A. Mukhopadhyay\\
       Baylor University\\
        E-mail: \email{aditi\_mukhopadhyay@baylor.edu}}
\author{ S.A. Yost%
     \thanks{Work supported in part by U.S.
D.o.E. grant DE-FG02-10ER41694 and grants from The Citadel Foundation.}\\
       The Citadel\\
        E-mail: \email{scott.yost@citadel.edu}}

\abstract{We present the interplay between the new IR-improved DGLAP-CS theory 
and the precision of NLO parton shower/ME matched MC`s as it is realized by 
the new MC Herwiri1.031 in interface to MC@NLO. We discuss phenomenological 
implications using comparisons with recent LHC data on single heavy gauge 
boson production.}

\FullConference{36th International Conference on High Energy Physics\\
                 4-11 July 2012\\
                 Melbourne, Australia\\
                 BU-HEPP-12-05, Nov., 2012}

\begin{document}
\baselineskip=10pt
\section{\bf Introduction}\label{intro}\par
With the recent announcement~\cite{atlas-cms-2012} 
of an Englert-Brout-Higgs (EBH)~\cite{EBH} candidate 
boson after the start-up and successful running 
of the LHC, the era of precision QCD, 
wherein the total precision tag is $1\%$ or better,
is upon us. The attendant need for exact, amplitude-based 
resummation of large higher order effects
is now more paramount, in view of the expected role of precision comparison between theory and experiment in determining the detailed properties of the newly discovered EBH boson candidate. It has been argued elsewhere~\cite{radcor2011,qced} that
such resummation allows one to have better than 1\% theoretical precision 
as a realistic goal in such comparisons. 
Here, we present the status of this approach to precision QCD for the LHC with the attendant IR-improved DGLAP-CS~\cite{dglap,cs} theory~\cite{irdglap1,irdglap2} realization via HERWIRI1.031~\cite{herwiri} 
in the HERWIG6.5~\cite{herwig} environment
in interplay with NLO exact, matrix element matched parton shower MC precision issues. We employ the MC@NLO~\cite{mcatnlo} methodology to realize
the attendant exact, NLO matrix element matched parton shower MC
realizations in comparisons with recent LHC data.
\par
In the discussion we continue the strategy of building on existing platforms to develop and realize a path toward precision QCD for the physics of the LHC. We exhibit a union of the new IR-improved DGLAP-CS theory and MC@NLO. We are also pursuing the implementation~\cite{elswh} of the new IR-improved 
DGLAP-CS theory for
HERWIG++~\cite{hwg++}, HERWIRI++,
for PYTHIA8~\cite{pyth8} and for SHERPA~\cite{shrpa}, as well as
the corresponding NLO ME/parton shower matching realizations in the POWHEG~\cite{pwhg} framework --
one of the strongest cross checks on theoretical precision is the difference between two independent realizations of the attendant theoretical calculation.
\par
We set the stage for the proper exposition of the interplay between the NLO ME matched parton shower MC precision and the new IR-improved DGLAP-CS theory in the next section by showing how the latter theory follows 
naturally in the effort to obtain
a provable precision from our approach~\cite{qced} to precision LHC physics. 
We review this latter approach in the next section as well. We then turn in Section 3 to the applications to the recent data on single heavy gauge boson production at the LHC with an eye on the analyses in Refs.~\cite{herwiri} of the analogous processes at the Tevatron. We will focus in this discussion on the single Z/$\gamma*$ production and decay to lepton pairs for definiteness. The other heavy gauge boson processes will be taken up elsewhere~\cite{elswh}. 
\par
\section{Brief Recapitulation}

The starting point for our discussion may be taken as the 
fully differential representation
\begin{equation}
d\sigma =\sum_{i,j}\int dx_1dx_2F_i(x_1)F_j(x_2)d\hat\sigma_{\text{res}}(x_1x_2s)
\label{bscfrla}
\end{equation}
of a hard LHC scattering process
using a standard notation so that the $\{F_j\}$ and 
$d\hat\sigma_{\text{res}}$ are the respective parton densities(PDFs) and 
reduced resummed hard differential cross section, where the resummation is 
for all large EW and QCD higher order corrections in order to
achieve a total precision tag of 1\% or better for the total 
theoretical precision of (\ref{bscfrla}). The proof of the correctness of the value of the 
total theoretical precision $\Delta\sigma_{\text{th}}$ of (\ref{bscfrla})
is the basic issue for precision
QCD for the LHC . 
This precision can be represented as follows:$\Delta\sigma_{\text{th}}= \Delta F \oplus\Delta\hat\sigma_{\text{res}}$
where $\Delta A$ is the contribution of the uncertainty
on the quantity $A$ to $\Delta\sigma_{\text{th}}$\footnote{Here, we discuss the situation in which the two errors in the equation for $\Delta\sigma_{\text{th}}$ are independent
for definiteness; the equation for it has to be modified accordingly when
they are not.}.
The
proof of the correctness of the value of the 
total theoretical precision $\Delta\sigma_{\text{th}}$
is essential for 
validation of the  application of a given 
theoretical prediction to precision 
experimental observations for the signals and the
backgrounds for 
both Standard Model(SM) and new physics (NP) studies, and more specifically
for the overall normalization
of the cross sections in such studies.
We cannot emphasize too much that NP can be missed if a calculation
with an unknown value of $\Delta\sigma_{\text{th}}$ 
is used for the attendant studies.
We note that here $\Delta\sigma_{\text{th}}$ is the 
total theoretical uncertainty that comes from the physical precision 
contribution and the technical precision contribution~\cite{jadach-prec}:
the physical precision contribution, $\Delta\sigma^{\text{phys}}_{\text{th}}$,
arises from such sources as missing graphs, approximations to graphs, 
truncations,....; the technical precision contribution, 
$\Delta\sigma^{\text{tech}}_{\text{th}}$, arises from such sources as 
bugs in codes, numerical rounding errors,
convergence issues, etc. The total theoretical error follows from 
\begin{equation}
\Delta\sigma_{\text{th}}=\Delta\sigma^{\text{phys}}_{\text{th}}\oplus \Delta\sigma^{\text{tech}}_{\text{th}}.
\end{equation}
As a general rule, the desired value for $\Delta\sigma_{\text{th}}$, which 
depends on the  specific
requirements of the observations, should fulfill
$\Delta\sigma_{\text{th}}\leq f\Delta\sigma_{\text{expt}}$. 
Here $\Delta\sigma_{\text{expt}}$ is the respective experimental error
and $f\lesssim \frac{1}{2}$ so that
the theoretical uncertainty does not significantly adversely affect the 
analysis of the data for physics studies.
\par
With the goal of realizing such precision in a provable way, we have 
developed the $\text{QCD}\otimes\text{QED}$ resummation theory in Refs.~\cite{qced}
for the reduced cross section in (\ref{bscfrla}) and for the
resummation of the evolution of the parton densities therein as well.
Mainly because the theory in
Refs.~\cite{qced} is not widely known, we recapitulate it here briefly.
The master formula for our resummation theory
may be identified as
\begin{eqnarray}
&d\bar\sigma_{\rm res} = e^{\rm SUM_{IR}(QCED)}
   \sum_{{n,m}=0}^\infty\frac{1}{n!m!}\int\prod_{j_1=1}^n\frac{d^3k_{j_1}}{k_{j_1}} \cr
&\prod_{j_2=1}^m\frac{d^3{k'}_{j_2}}{{k'}_{j_2}}
\int\frac{d^4y}{(2\pi)^4}e^{iy\cdot(p_1+q_1-p_2-q_2-\sum k_{j_1}-\sum {k'}_{j_2})+
D_\rQCED} \cr
&\tilde{\bar\beta}_{n,m}(k_1,\ldots,k_n;k'_1,\ldots,k'_m)\frac{d^3p_2}{p_2^{\,0}}\frac{d^3q_2}{q_2^{\,0}},
\label{subp15b}
\end{eqnarray}\noindent
where $d\bar\sigma_{\rm res}$ is either the reduced cross section
$d\hat\sigma_{\rm res}$ or the differential rate associated to a
DGLAP-CS~\cite{dglap,cs} kernel involved in the evolution of the $\{F_j\}$ and 
where the {\em new} (YFS-style~\cite{yfs}) {\em non-Abelian} residuals 
$\tilde{\bar\beta}_{n,m}(k_1,\ldots,k_n;k'_1,\ldots,k'_m)$ have $n$ hard gluons and $m$ hard photons and we show the final state with two hard final
partons with momenta $p_2,\; q_2$ specified for a generic $2f$ final state for
definiteness. The infrared functions ${\rm SUM_{IR}(QCED)},\; D_\rQCED\; $
are defined in Refs.~\cite{qced,irdglap1,irdglap2}. This  
simultaneous resummation of QED and QCD large IR effects is exact. 
\par
We have shown in Refs.~\cite{irdglap1,irdglap2,herwiri}
that the methods in Refs.~\cite{stercattrent1,scet1}
give approximations to our hard gluon residuals  $\hat{\tilde{\bar\beta}}_{n}$;
 for, the methods in Refs.~\cite{stercattrent1,scet1}, unlike the master formula in (\ref{subp15b}), are not exact results. The threshold-resummation
methods in Refs.~\cite{stercattrent1}, using the result
that, for any function $f(z)$,
$$\left|\int_0^1 dz z^{n-1}f(z)\right|\le(\frac{1}{n})\max_{z\in [0,1]} {|f(z)|},$$
drop non-singular contributions to the cross section at $z\rightarrow 1$
in resumming the logs in $n$-Mellin space. The SCET theory in Refs.~\cite{scet1}
drops terms of ${\cal O}(\lambda)$ at the level of the amplitude, where $\lambda=\sqrt{\Lambda/Q}$ for a process with the hard scale $Q$ with $\Lambda \sim .3\text{GeV}$ so that, for $Q\sim 100\text{GeV}$, $\lambda\cong 5.5\%$.
The known equivalence of the two approaches implies that the errors in the threshold resummation must be similar. We can only use these approaches as a guide to our new non-Abelian residuals as we develop results for the sub-1\% precision regime.\par
As it is explained in Refs.~\cite{qced}, 
the new non-Abelian residuals $\tilde{\bar\beta}_{m,n}$ 
allow rigorous shower/ME matching via their shower subtracted analogs:
$\tilde{\bar\beta}_{m,n}\rightarrow \hat{\tilde{\bar\beta}}_{m,n}$
where the $\hat{\tilde{\bar\beta}}_{m,n}$ have had all effects in the showers
associated to the $\{F_j\}$ removed from them and 
this naturally brings us to the attendant evolution of the $\{F_j\}$.  
For a strict control on the theoretical precision
in (\ref{bscfrla}), we need both the resummation of the reduced cross section
and that of the latter evolution.
\par 
When the QCD restriction of the formula in (\ref{subp15b}) is applied to the
calculation of the kernels, $P_{AB}$, in the DGLAP-CS theory itself, 
we get an improvement
of the IR limit of these kernels, an IR-improved DGLAP-CS theory~\cite{irdglap1,irdglap2,herwiri} with
new resummed kernels  $P^{\exp}_{AB}$, which are reproduced here for completeness:
{\small
\begin{align}
P^{\exp}_{qq}(z)&= C_F \FYFS(\gamma_q)e^{\frac{1}{2}\delta_q}\left[\frac{1+z^2}{1-z}(1-z)^{\gamma_q} -f_q(\gamma_q)\delta(1-z)\right],\nonumber\\
P^{\exp}_{Gq}(z)&= C_F \FYFS(\gamma_q)e^{\frac{1}{2}\delta_q}\frac{1+(1-z)^2}{z} z^{\gamma_q},\nonumber\\
P^{\exp}_{GG}(z)&= 2C_G \FYFS(\gamma_G)e^{\frac{1}{2}\delta_G}\{ \frac{1-z}{z}z^{\gamma_G}+\frac{z}{1-z}(1-z)^{\gamma_G}\nonumber\\
&\qquad +\frac{1}{2}(z^{1+\gamma_G}(1-z)+z(1-z)^{1+\gamma_G}) - f_G(\gamma_G) \delta(1-z)\},\nonumber\\
P^{\exp}_{qG}(z)&= \FYFS(\gamma_G)e^{\frac{1}{2}\delta_G}\frac{1}{2}\{ z^2(1-z)^{\gamma_G}+(1-z)^2z^{\gamma_G}\},
\label{dglap19}
\end{align}}
where the superscript ``$\exp$'' indicates that the kernel has been resummed as
predicted by Eq.\ (\ref{subp15b}) when it is restricted to QCD alone, where
the YFS~\cite{yfs} infrared factor 
is given by $\FYFS(a)=e^{-C_Ea}/\Gamma(1+a)$ where $C_E$ is Euler's constant
and where the respective resummation functions $\gamma_A,\;\delta_A,\; f_A,\; A\; =\; q, \; G$
are given in Refs.~\cite{irdglap1,irdglap2}
\footnote{The improvement in Eq.\ (\ref{dglap19}) 
should be distinguished from the 
resummation in parton density evolution for the ``$z\rightarrow 0$'' 
Regge regime -- see for example Ref.~\cite{ermlv,guido}. This
latter improvement must also be taken into account 
for precision LHC predictions.}. $C_F$($C_G$) is the quadratic Casimir invariant for the quark(gluon) color representation respectively.
From these new kernels we get
a new resummed scheme for the PDFs and the reduced cross section: 
\begin{equation}
\begin{split}
F_j,\; \hat\sigma &\rightarrow F'_j,\; \hat\sigma'\; \text{for}\\
P_{Gq}(z)&\rightarrow P^{\exp}_{Gq}(z), \text{etc.},
\end{split}
\label{newscheme1}
\end{equation}
with the same value for $\sigma$ in (\ref{bscfrla}) with improved MC stability~\cite{herwiri} 
-- we do not need 
an IR cut-off `$k_0$'
parameter in the attendant parton shower MC based on the new kernels.
Note that, while the degrees of freedom
below the IR cut-offs in the usual showers are dropped in those showers,
in the showers in HERWIRI1.031,
as one can see from (\ref{subp15b}), these degrees of freedom are integrated over and included in the calculation in the process of generating the Gribov-Lipatov exponents $\gamma_A$ in (\ref{dglap19}). The new kernels
agree with the usual kernels at ${\cal O}(\alpha_s)$ as the differences between them start in ${\cal O}(\alpha_s^2)$, so that the NLO matching formulas
in the MC@NLO and POWHEG frameworks apply directly 
to the new kernels for exact
NLO ME/shower matching. 
\par
In Fig.~1 we show the basic physical idea of Bloch and Nordsieck~\cite{bn1}
underlying the new kernels: 
\begin{figure}[h]
\begin{center}
\epsfig{file=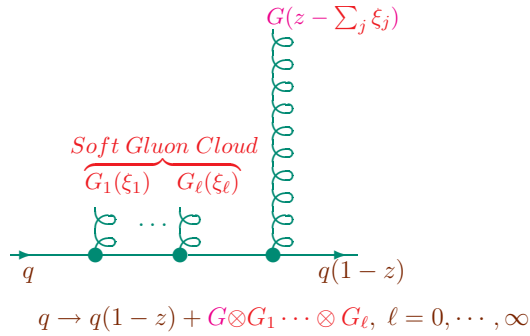,width=70mm}
\end{center}
\label{fig-bn-1}
\caption{Bloch-Nordsieck soft quanta for an accelerated charge.}
\end{figure}
an accelerated charge generates a coherent state of very soft massless quanta of the respective gauge field so that one cannot know which of the infinity of possible states
one has made in the splitting process $q(1)\rightarrow q(1-z)+G\otimes G_1\cdots\otimes G_\ell,\; \ell=0,\cdots,\infty$ illustrated in Fig.~1.
The new kernels take this effect into account by resumming the 
terms ${\cal O}\left((\alpha_s \ln(\frac{q^2}{\Lambda^2})\ln(1-z))^n\right)$
when $z\rightarrow 1$ is the IR limit. From (\ref{newscheme1}) and (\ref{bscfrla}), one can see that
when the usual kernels are used these terms
are generated order-by-order in the solution for the cross section
$\sigma$ in (\ref{bscfrla}). Our 
resumming of these terms enhances the convergence of the 
representation in (\ref{bscfrla}) for a given order of exactness in the
in input perturbative components therein.  In the next section we
illustrate this last remark in the context of the comparison of 
NLO parton shower/matrix element matched predictions to recent LHC data.\par

\section{Interplay of NLO Shower/ME Precision and IR-Improved DGLAP-CS Theory}

Here, 
we compare new MC HERWIRI1.031~\cite{herwiri} with HERWIG6.510, both with and without
the MC@NLO~\cite{mcatnlo} exact ${\cal O}(\alpha_s)$ correction
to illustrate the interplay between the attendant precision in NLO ME matched parton shower MC's  
and the new IR-improvement for the kernels realized in Herwiri1.031, where we use the new LHC data for our baseline for the comparison.\par
\begin{figure}[h]
\begin{center}
\setlength{\unitlength}{0.1mm}
\begin{picture}(1600, 930)
\put( 370, 770){\makebox(0,0)[cb]{\bf (a)} }
\put(1240, 770){\makebox(0,0)[cb]{\bf (b)} }
\put(   -50, 0){\makebox(0,0)[lb]{\includegraphics[width=80mm]{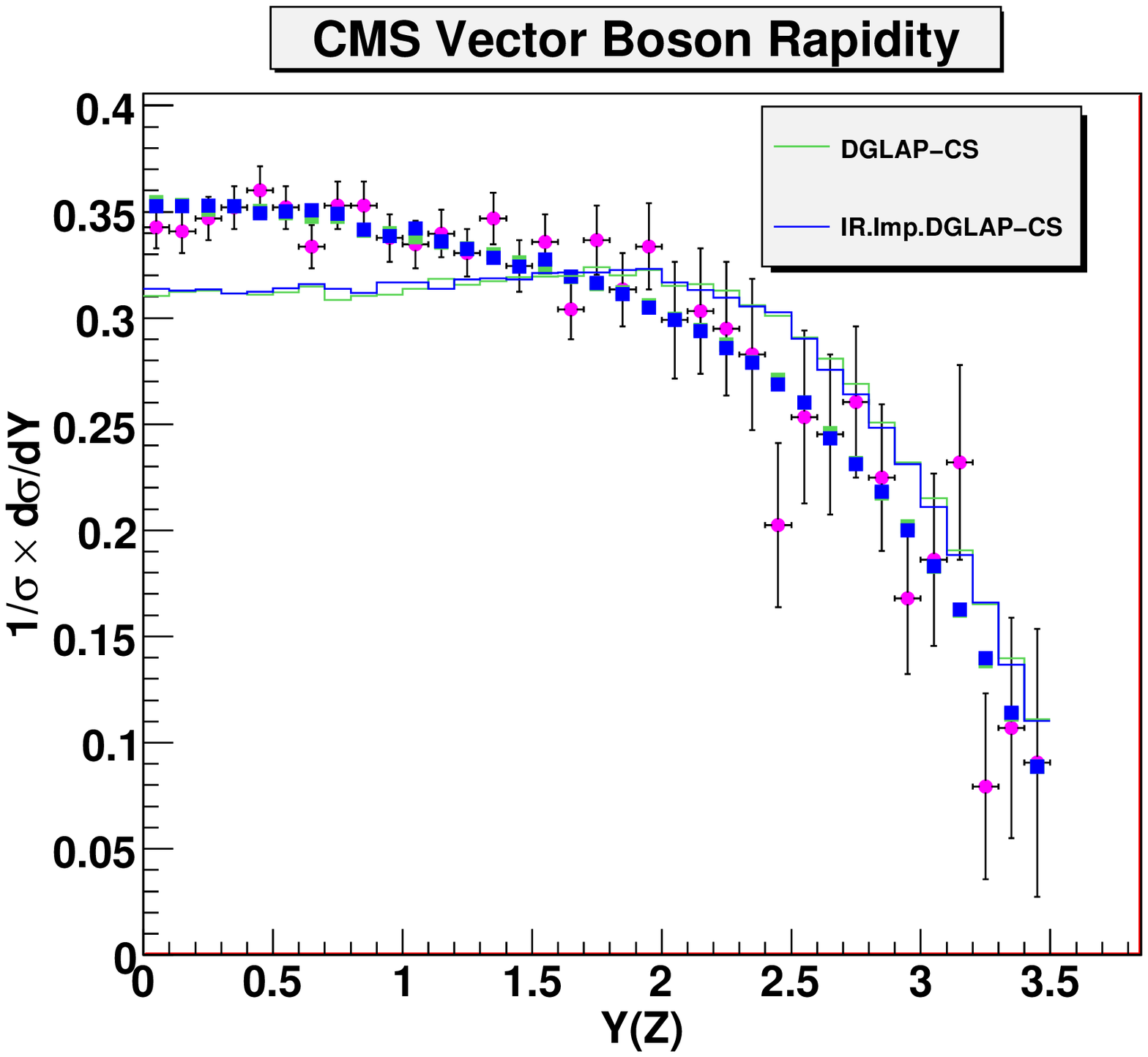}}}
\put( 830, 0){\makebox(0,0)[lb]{\includegraphics[width=80mm]{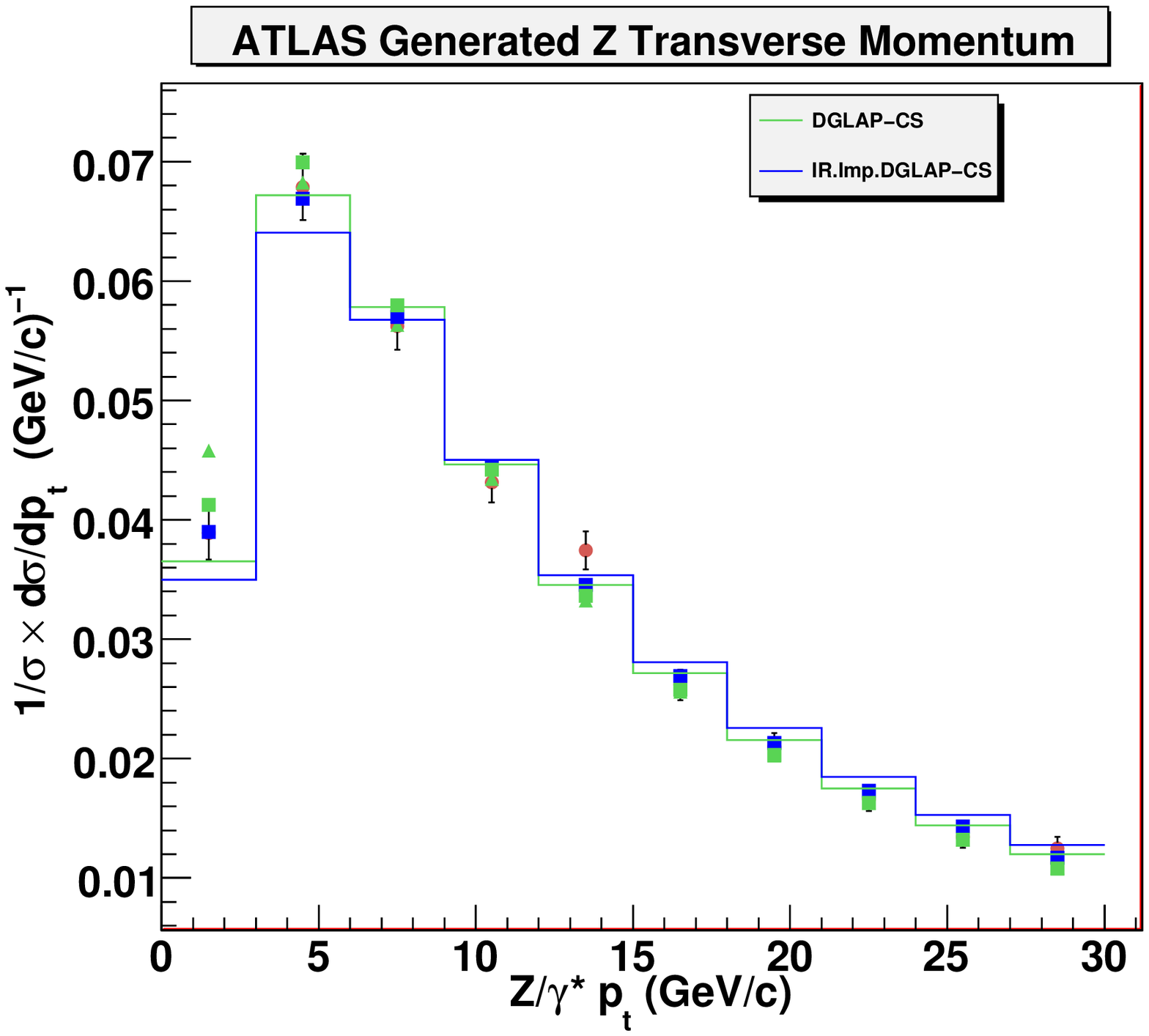}}}
\end{picture}
\end{center}
\caption{\baselineskip=8pt Comparison with LHC data: (a), CMS rapidity data on
($Z/\gamma^*$) production to $e^+e^-,\;\mu^+\mu^-$ pairs, the circular dots are the data, the green(blue) lines are HERWIG6.510(HERWIRI1.031); 
(b), ATLAS $p_T$ spectrum data on ($Z/\gamma^*$) production to (bare) $e^+e^-$ pairs,
the circular dots are the data, the blue(green) lines are HERWIRI1.031(HERWIG6.510). In both (a) and (b) the blue(green) squares are MC@NLO/HERWIRI1.031(HERWIG6.510($\rm{PTRMS}=2.2$GeV)). In (b), the green triangles are MC@NLO/HERWIG6.510($\rm{PTRMS}=$0). These are otherwise untuned theoretical results. 
}
\label{fig2-nlo-iri}
\end{figure}

For the single $Z/\gamma*$ production at the LHC, we show in Fig.~\ref{fig2-nlo-iri} in panel (a) the comparison between the MC predictions and the CMS rapidity data~\cite{cmsrap} and in panel
(b) the analogous comparison with the ATLAS $P_T$ data, where the rapidity data  are the combined $e^+e^--\mu^-\mu^+$ results and the $p_T$ data are those for the bare $e^+e^-$ case, as these are the data that correspond to the theoretical
framework of our simulations -- we do not as yet have complete realization of all the corrections involved in the other ATLAS data in Ref.~\cite{atlaspt}. 
These results should be viewed with an eye on our analysis in Ref.~\cite{herwiri} of the FNAL data on the single $Z/\gamma^*$ production in 
$\text{p}\bar{\text{p}}$ collisions at 1.96 TeV.\par
In Fig.~11 of the second paper in Ref.~\cite{herwiri}, we showed that, when the intrinsic rms $p_T$ parameter $\rm{PTRMS}$ is set to 0 in HERWIG6.5, the simulations for MC@NLO/HERWIG6.510 give a good fit to the CDF rapidity distribution data~\cite{galea} therein but they do not give a satisfactory fit to the D0 $p_T$ distribution data~\cite{d0pt} therein whereas the results for MC@NLO/HERWIRI1.031 give good fits to both sets of data with the $\rm{PTRMS} =0$. Here $\rm{PTRMS}$ corresponds to an intrinsic Gaussian distribution in $p_T$. The authors of HERWIG~\cite{mike2} have emphasized that to get good fits to both sets of data, one may set $\rm{PTRMS}\cong 2$ GeV. Thus, in analyzing the new LHC data, we have set $\rm{PTRMS}=2.2$GeV in our HERWIG6.510 simulations while we continue to set $PTRMS=0$ in our HERWIRI simulations.
\par
Turning now with this perspective to the results in Fig.~\ref{fig2-nlo-iri}, we see confirmation of the finding of the HERWIG authors. To get a good fit to both the CMS rapidity data and the ATLAS $p_T$ data, one needs to set $\rm{PTRMS}\cong 2 \text{GeV}$~\cite{skands} in the MC@NLO/HERWIG6510 simulations. We again see that at LHC one gets a good fit to the data for both the rapidity and the $p_T$ spectra in the MC@NLO/HERWIRI1.031 simulations with $\rm{PTRMS}=0$. In quantitative terms, the $\chi^2/\text{d.o.f.}$ for the rapidity data and $p_T$ data are (.72,.72)((.70,1.37)) for the 
MC@NLO/HERWIRI1.031(MC@NL
O/HERWIG
6510($\rm{PTRMS}$=2.2GeV)) simulations. For the
 MC@NLO/HERWIG6510($\rm{PTRMS}$=0) simulations the corresponding results are (.70,2.23).
\par 
The usual DGLAP-CS kernels require the introduction of a hard intrinsic Gaussian spread in $p_T$  inside the proton to reproduce the LHC data on the $p_T$ distribution of the 
$Z/\gamma^*$ in the pp collisions whereas the IR-improved kernels give a better fit to the data without the introduction of such. This hard $\rm{PTRMS}$ is entirely ad hoc; it is in contradiction with the results of all successful models of the proton wave-function~\cite{pwvfn}, wherein the scale of it is $\lesssim .4$GeV. More importantly, it contradicts the known experimental observation of precocious Bjorken scaling~\cite{scaling,bj1}, where the SLAC-MIT experiments show that Bjorken scaling occurs already at $Q^2=1_+$ GeV$^2$
for $Q^2=-q^2$ with q the 4-momentum transfer from the electron to the proton
in the famous deep inelastic electron-proton scattering process whereas, if the proton constituents really had a Gaussian intrinsic $p_T$ distribution with $\rm{PTRMS}\cong 2$GeV, these observations 
would not be possible. What can now argue is that the ad hoc $\rm{PTRMS}\cong 2.2$GeV value is a phenomenological representation of the more fundamental dynamics realized by the IR-improved DGLAP-CS theory. Is it possible
to tell the difference between the two representations of the data in 
Fig.~\ref{fig2-nlo-iri}?\par
\begin{figure}[h]
\begin{center}
\epsfig{file=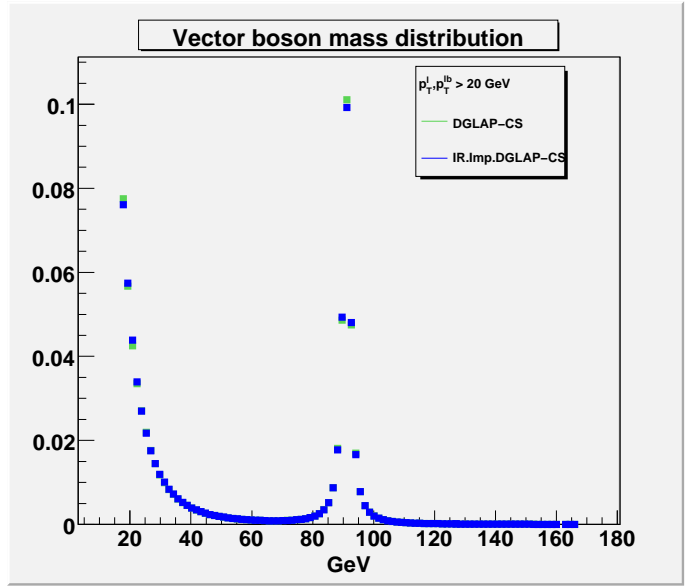,width=90mm}
\end{center}
\label{fig3-nlo-iri}
\caption{Normalized vector boson mass spectrum at the LHC for $p_T(\text{lepton}) >20$ GeV.}
\end{figure}

Physically, one expects that more detailed observations should be able to distinguish the two. Indeed, in Fig.~\ref{fig3-nlo-iri} we show the  MC@NLO/HERWIRI
1.031(blue squares) and MC@NLO/HER
WIG6510($\rm{PTRMS}$=2.2GeV) (green squares) predictions
for the $Z/\gamma^*$ mass spectrum when the decay lepton pairs are required to 
satisfy the LHC type requirement that their transverse momenta $\{p^\ell_T, p^{\bar\ell}_T\}$ exceed $20$ GeV.
As the peaks differ by 2.2\%, the 
high precision data such as the LHC ATLAS and CMS
experiments will have
(each already has over $5\times 10^6$ lepton pairs) will allow one to distinguish between the two sets
of theoretical predictions. Other such detailed observations
may also reveal 
the differences between the two representations of parton shower physics
and we will pursue these elsewhere~\cite{elswh}.
In closing, two of us (A.M. and B.F.L.W.)
thank Prof. Ignatios Antoniadis for the support and kind 
hospitality of the CERN TH Unit while part of this work was completed.\par


\begin{thebibliography}{99}
\bibitem{atlas-cms-2012} F. Gianotti, in {\it Proc. ICHEP2012}, in press; J. Incandela, {\it ibid.}, 2012, in press; G. Aad {\it et al.}, arXiv:1207.7214; D. Abbaneo {\it et al.}, arXiv:1207.7235.
\bibitem{EBH} F. Englert and R. Brout, Phys. Rev. Lett. {\bf 13} (1964) 312; P.W. Higgs, Phys. Lett. {\bf 12} (1964) 132; Phys. Rev. Lett. {\bf 13} (1964) 508;
G.S. Guralnik, C.R. Hagen and T.W.B. Kibble, {\it ibid.} {\bf 13} (1964) 585.
\bibitem{radcor2011} B.F.L. Ward, S.K. Majhi and S.A. Yost, in PoS({\bf RADCOR2011}) (2012) 022. 
\bibitem{qced} C. Glosser, S. Jadach, B.F.L. Ward and S.A. Yost, Mod. Phys. Lett. A {\bf 19}(2004) 2113; B.F.L. Ward, C. Glosser, S. Jadach and S.A. Yost, in {\it Proc. DPF 2004}, Int. J. Mod. Phys. A {\bf 20} (2005) 3735; in {\it Proc. ICHEP04, vol. 1}, eds. H. Chen {\it et al.},(World. Sci. Publ. Co., Singapore, 2005) p. 588; B.F.L. Ward and S. Yost, preprint BU-HEPP-05-05, in {\it Proc. HERA-LHC Workshop}, CERN-2005-014; in {\it  Moscow 2006, ICHEP, vol. 1}, p. 505; Acta Phys. Polon. B {\bf 38} (2007) 2395; arXiv:0802.0724, PoS{\bf(RADCOR2007)}(2007) 038; B.F.L. Ward {\it et al.}, arXiv:0810.0723, in {\it Proc. ICHEP08}; arXiv:0808.3133, in {\it Proc. 2008 HERA-LHC Workshop},DESY-PROC-2009-02, eds. H. Jung and A. De Roeck, (DESY, Hamburg, 2009) p. 168, and references therein.
\bibitem{dglap}
G. Altarelli and G. Parisi, {\it Nucl. Phys.} {\bf B126} (1977) 
298; Yu. L. Dokshitzer, {\it Sov. Phys. JETP} {\bf 46} (1977) 641;
L.~N. Lipatov, {\it Yad. Fiz.} {\bf 20} (1974) 181; V. Gribov and L. Lipatov,
{\it Sov. J. Nucl. Phys.} {\bf 15} (1972) 675, 938; see also J.C. Collins and J. Qiu,
{\it Phys. Rev. D}{\bf 39} (1989) 1398. 
\bibitem{cs}C.G. Callan, Jr., {\it Phys. Rev. D}{\bf 2} (1970) 1541; K. Symanzik, 
{\it Commun. Math. Phys.} {\bf 18} (1970) 227, 
and in {\em Springer Tracts in Modern Physics},
{\bf 57}, ed. G. Hoehler (Springer, Berlin, 1971) p. 222; see also
S. Weinberg, {\it Phys. Rev. D}{\bf 8} (1973) 3497.
\bibitem{irdglap1} B.F.L. Ward, {\it Adv. High Energy Phys.} {\bf 2008} (2008) 682312. 
\bibitem{irdglap2} B.F.L. Ward, {\it Ann. Phys.} {\bf 323} (2008) 2147.
\bibitem{herwiri} S. Joseph {\it et al.}, Phys. Lett. B{\bf 685} (2010) 283; Phys. Rev. D{\bf 81} (2010) 076008.
\bibitem{herwig} G. Corcella {\it et al.}, hep-ph/0210213; 
J. High Energy Phys. {\bf 0101} (2001) 010; 
G. Marchesini {\it et al.}, Comput. Phys. Commun.{\bf 67} (1992) 465. 
\bibitem{mcatnlo} S. Frixione and B.Webber, J. High Energy Phys. {\bf 0206} (2002) 029; S. Frixione {\it et al.}, arXiv:1010.0568.
\bibitem{elswh} A. Mukhopadhyay {\it et al.}, to appear.
\bibitem{hwg++} M. Bahr {\it et al.}, arXiv:0812.0529 and references therein.
\bibitem{pyth8} T. Sjostrand, S. Mrenna and P. Z. Skands, Comput. Phys. Commun. {\bf 178} (2008) 852-867.
\bibitem{shrpa} T. Gleisberg {\it et al.}, J.High Energy Phys. {\bf 0902}
(2009) 007. 
\bibitem{pwhg} P. Nason, J. High Energy Phys. {\bf 0411} (2004) 040.
\bibitem{jadach-prec} See for example S. Jadach {\it et al.}, in {\it Physics at LEP2, vol. 2}, (CERN, Geneva, 1995) pp. 229-298.
\bibitem{yfs} D. R. Yennie, S. C. Frautschi, and H. Suura, Ann. Phys. {\bf 13} (1961) 379; see also K. T. Mahanthappa, Phys. Rev. {\bf 126} (1962) 329, for a related analysis.
\bibitem{stercattrent1} G. Sterman,{\it Nucl. Phys.} {\bf B281}, 310 (1987); S. Catani and L. Trentadue,
{\it Nucl. Phys.} {\bf B327}, 323 (1989); {\it ibid.} {\bf B353}, 183 (1991).
\bibitem{scet1} See for example C. W. Bauer, A.V. Manohar and M.B. Wise, {\it Phys. Rev. Lett.} {\bf 91} (2003) 122001; {\it Phys. Rev.} {\bf D70} (2004) 034014; C. Lee and G. Sterman, {\it Phys. Rev. D} {\bf 75} (2007) 014022.
\bibitem{ermlv} B.I. Ermolaev, M. Greco and S.I. Troyan, {\it PoS DIFF2006} (2006) 036, and references therein.
\bibitem{guido} G. Altarelli, R.D. Ball and S. Forte, 
{\it PoS RADCOR2007} (2007) 028.
\bibitem{bn1} F. Bloch and A. Nordsieck, Phys. Rev. {\bf 52} (1937) 54.
\bibitem{madg} S.M. Abyat {\it et al.}, {\it Phys. Rev. D} {\bf 74} (2006) 074004.
\bibitem{cmsrap} S. Chatrchyan {\it et al.}, arXiv:1110.4973; Phys. Rev. D{\bf 85} (2012) 032002.
\bibitem{atlaspt} G. Aad {\it et al.}, arXiv:1107.2381; Phys. Lett. B{\bf 705} (2011) 415.
\bibitem{mike2} M. Seymour, ``Event Generator Physics for the LHC'', CERN Seminar, 2011.
\bibitem{galea} C. Galea, in {\it Proc. DIS 2008}, London, 2008,\newline 
\verb$http://dx.doi.org/10.3360/dis.2008.55$.
\bibitem{d0pt} V.M. Abasov {\it et al.}, {\it Phys. Rev. Lett.} {\bf 100}, 102002 (2008).
\bibitem{skands} P. Skands, private communication, 2011, finds a similar behavior in PYTHIA8 simulations.
\bibitem{pwvfn} R.P. Feynman, M. Kislinger and F. Ravndal, Phys. Rev. D{\bf 3} (1971) 2706; R. Lipes, {\it ibid.}{\bf 5} (1972) 2849; F.K. Diakonas, N.K. Kaplis and X.N. Mawita, {\it ibid.} {\bf 78} (2008) 054023; K. Johnson, {\it Proc. Scottish Summer School Phys. 17} (1976) p. 245; A. Chodos et al., Phys. Rev. D{\bf 9} (1974) 3471; {\it ibid.} {\bf 10} (1974) 2599; T. DeGrand {\it et al.}, {\it ibid.} {\bf 12} (1975) 2060.
\bibitem{scaling} See for example R.E. Taylor, Phil. Trans. Roc. Soc. Lond. {\bf A359} (2001) 225, and references therein.
\bibitem{bj1} J. Bjorken, in {\it Proc. 3rd International Symposium on the History of Particle Physics: The Rise of the Standard Model, Stanford, CA, 1992}, eds. L. Hoddeson {\it et al.} (Cambridge Univ. Press, Cambridge, 1997) p. 589, and references therein.
\end{thebibliography}
\end{document}